\documentclass[conference]{IEEEtran}
\IEEEoverridecommandlockouts
\usepackage{listings}
\usepackage{color}

\definecolor{lightgray}{gray}{0.9}

\newcommand{\inlinecode}[2]{\colorbox{lightgray}{\lstinline[language=#1]$#2$}}

\usepackage{enumitem,kantlipsum} 
\usepackage{placeins}
\usepackage{makecell, multirow}
\usepackage{mwe}
\usepackage{subcaption}
\usepackage{multicol}
\usepackage{booktabs,tabularx}
\usepackage{amsmath,amssymb,amsfonts}
\usepackage{float}
\usepackage[maxbibnames=99]{biblatex}
\usepackage[utf8]{inputenc} 
\usepackage{fancyhdr}
\usepackage{blindtext}
\usepackage{csquotes}
\usepackage[misc]{ifsym}  
\usepackage{braket}
\usepackage[
  separate-uncertainty = true,
  multi-part-units = repeat
]{siunitx}
\usepackage{url}
\usepackage{hyperref}
\usepackage{cleveref }
\usepackage{algorithmic}
\usepackage{algorithm}
\usepackage{graphicx}
\usepackage{textcomp}
\usepackage{xcolor}    
\usepackage{caption}
\usepackage{tikz}
\usepackage{pgfplots, pgfplotstable}
\pgfplotsset{compat=1.17}

\usepackage{tabularx,booktabs}
\newcolumntype{C}{>{\centering\arraybackslash}X} % centered version of "X" type
\usepackage{lipsum}
\def\BibTeX{{\rm B\kern-.05em{\sc i\kern-.025em b}\kern-.08em
    T\kern-.1667em\lower.7ex\hbox{E}\kern-.125emX}}
     
\begin{document}

\title{Depression detection from Social Media Bangla Text  Using Recurrent Neural Networks}

\author{
\IEEEauthorblockN{Sultan Ahmed}
\IEEEauthorblockA{\textit{Department of Information Systems} \\
\textit{University of Maryland Baltimore County}\\
Maryland, USA \\
IL66977@umbc.edu}

\and 

\IEEEauthorblockN{Salman Rakin}
\IEEEauthorblockA{\textit{Department of CSE} \\
\textit{Bangladesh University }\\
\textit{ of Engineering \& Technology}\\
Dhaka, Bangladesh \\
0417052033@grad.cse.buet.ac.bd}

\and 

\IEEEauthorblockN{Mohammad Washeef Ibn Waliur}
\IEEEauthorblockA{\textit{Department of CSE} \\
\textit{University of Dhaka}\\
Dhaka, Bangladesh \\
washeef123@gmail.com}

\and 
\IEEEauthorblockN{Nuzhat Binte Islam}
\IEEEauthorblockA{\textit{Department of URP} \\
\textit{Jahangirnagar University}\\
Dhaka, Bangladesh \\
mounuzhat35@gmail.com}

\and 
\IEEEauthorblockN{Billal Hossain}
\IEEEauthorblockA{\textit{Department of CSE} \\
\textit{Jagannath University}\\
Dhaka, Bangladesh \\
billal13027@gmail.com}

 \and 
\IEEEauthorblockN{Dr. Md. Mostofa Akbar}
\IEEEauthorblockA{\textit{Department of CSE} \\ 
\textit{Bangladesh University }\\
\textit{ of Engineering \& Technology}\\
Dhaka, Bangladesh \\
mostofa@cse.buet.ac.bd}
}

\maketitle
%%%%%%%%%%%%%%%%%%%%%%%%%%%%%%%%%%%%%%%%%%%%%%%%%%%%%%%%%%%%%%%%%%%%%%%%%
% This section is based on the bbk10.clo file
% of Palash Baran Pal's bangtex
% http://www.saha.ac.in/theory/palashbaran.pal/bangtex/bangtex.html
%%%%%%%%%%%%%%%%%%%%%%%%%%%%%%%%%%%%%%%%%%%%%%%%%%%%%%%%%%%%%%%%%%%%%%%%%

\def\sbng{\bngviii}
\def\tbng{\bngvi}
\def\bng{\bngx}
\def\lbng{\bngxiv}
\def\Lbng{\bngxviii}
\def\LBng{\bngxxii}
\def\hbng{\bngxxv}
\def\Hbng{\bngxxx}
\def\sbns{\bnsviii}
\def\tbns{\bnsvi}
\def\bns{\bnsx}
\def\lbns{\bnsxiv}
\def\Lbns{\bnsxviii}
\def\LBns{\bnsxxii}
\def\hbns{\bnsxxv}
\def\Hbns{\bnsxxx}
\def\sbnw{\bnwviii}
\def\tbnw{\bnwvi}
\def\bnw{\bnwx}
\def\lbnw{\bnwxiv}
\def\Lbnw{\bnwxviii}
\def\LBnw{\bnwxxii}
\def\hbnw{\bnwxxv}
\def\Hbnw{\bnwxxx}

%%%%%%%%%%%%%%%%%%%%%%%%%%%%%%%%%%%%%%%%%%%%%%%%%%%%%%%%%%%%%%%%%%%%%%%%%
% This section is based on the bangfont.tex file
% of Palash Baran Pal's bangtex
% http://www.saha.ac.in/theory/palashbaran.pal/bangtex/bangtex.html
%%%%%%%%%%%%%%%%%%%%%%%%%%%%%%%%%%%%%%%%%%%%%%%%%%%%%%%%%%%%%%%%%%%%%%%%%

%%
%% Defining the normal bangla fornts
%%

\font\bngv=bang10 scaled 500
\font\bngvi=bang10 scaled 600
\font\bngvii=bang10 scaled 700
\font\bngviii=bang10 scaled 800
\font\bngix=bang10 scaled 900
\font\bngx=bang10
\font\bngxi=bang10 scaled 1100
\font\bngxii=bang10 scaled 1200
\font\bngxiv=bang10 scaled 1400
\font\bngxviii=bang10 scaled 1800
\font\bngxxii=bang10 scaled 2200
\font\bngxxv=bang10 scaled 2500
\font\bngxxx=bang10 scaled 3000

%%
%% Defining the slanted bangla fonts
%%
\font\bnsv=bangsl10 scaled 500
\font\bnsvi=bangsl10 scaled 600
\font\bnsvii=bangsl10 scaled 700
\font\bnsviii=bangsl10 scaled 800
\font\bnsix=bangsl10 scaled 900
\font\bnsx=bangsl10
\font\bnsxi=bangsl10 scaled 1100
\font\bnsxii=bangsl10 scaled 1200
\font\bnsxiv=bangsl10 scaled 1400
\font\bnsxviii=bangsl10 scaled 1800
\font\bnsxxii=bangsl10 scaled 2200
\font\bnsxxv=bangsl10 scaled 2500
\font\bnsxxx=bangsl10 scaled 3000

%%
%% Defining the wide bangla fonts
%%
\font\bnwv=bangwd10 scaled 500
\font\bnwvi=bangwd10 scaled 600
\font\bnwvii=bangwd10 scaled 700
\font\bnwviii=bangwd10 scaled 800
\font\bnwix=bangwd10 scaled 900
\font\bnwx=bangwd10
\font\bnwxi=bangwd10 scaled 1100
\font\bnwxii=bangwd10 scaled 1200
\font\bnwxiv=bangwd10 scaled 1400
\font\bnwxviii=bangwd10 scaled 1800
\font\bnwxxii=bangwd10 scaled 2200
\font\bnwxxv=bangwd10 scaled 2500
\font\bnwxxx=bangwd10 scaled 3000

%%
%% Inhibiting linebreak within words
%%
%\hyphenpenalty=10000 \pretolerance=-1 \tolerance=10000

%%
%% Defining the macro for e-kar, i-kar etc
%%
\def\*#1*#2{o\null{#2}{#1}}

%%
%% Redefining some macros to make them consistent with bangla fonts
%%
\def\d#1{\oalign{\smash{#1}\crcr\hidewidth{$\!$\rm.}\hidewidth}}

%%
%% Emulating the bold font
%%
\def\sh#1{\setbox0=\hbox{#1}%
     \kern-.02em\copy0\kern-\wd0
     \kern.04em\copy0\kern-\wd0
     \kern-.02em\raise.0433em\box0 }

\begin{abstract}

% 5\textsuperscript{th}
 Emotion artificial intelligence is a field of study that focuses on figuring out how to recognize emotions, especially in the area of text mining. Today is the age of social media which has opened a door for us to share our individual expressions, emotions, and perspectives on any event. We can analyze sentiment on social media posts to detect positive, negative, or emotional behavior toward society. One of the key challenges in sentiment analysis is to identify depressed text from social media text that is a root cause of mental ill-health.  Furthermore, depression leads to severe impairment in day-to-day living and is a major source of suicide incidents. In this paper, we apply natural language processing techniques on Facebook texts for conducting emotion analysis focusing on depression using multiple machine learning algorithms. Preprocessing steps like stemming, stop word removal, etc. are used to clean the collected data, and feature extraction techniques like stylometric feature, TF-IDF, word embedding, etc. are applied to the collected dataset which consists of 983 texts collected from social media posts. In the process of class prediction, LSTM, GRU, support vector machine, and Naive-Bayes classifiers have been used. We have presented  the results using the primary classification metrics including F1-score, and accuracy. This work focuses on depression detection from social media posts to help psychologists to analyze sentiment from shared posts which may reduce the undesirable behaviors of depressed individuals through diagnosis and treatment.    
\end{abstract}


\section{Introduction}
Text is the most important means of communication in today's world. Popular online social networking sites such as Facebook, Twitter, MySpace, etc. are mainly text-based. The rapid growth of Social Media has created enough opportunities to share information across time and space. Users are now comfortable contributing more to the content of social media websites and posting their own material. 

The emergence of internet-based media sources has resulted in the availability of substantial user data for the emotional analysis of text and images. Through a multitude of social media platforms such as Twitter, Facebook, and Instagram, individuals tend to express their emotions, opinions, and daily lives~\cite{Rambocas_Gama} today. This expression can be conveyed through images, videos, and text primarily. Due to the pervasiveness and accessibility of these social media platforms, there is an abundance of user data available for exploratory analysis. Textual data, as the most prevalent form of communication, possesses a number of characteristics that make it the optimal data source for emotion AI data analysis. 

People are increasingly utilizing social media to share and express their happiness, anxiety, sorrow, curiosity, etc. through textual content. Therefore, social media has emerged as a data source for analyzing people's expressions~\cite{hassan2017sentiment}. The majority of individuals in our country use Facebook to express their emotions. If an automated system is able to identify the depressive language in social media posts, it will be possible to identify depressed individuals within our network before they enter a malignant phase of depression.
   
The classical approach to feature extraction for depression analysis from textual data is to identify unique stylometric
features of written texts. The underlying assumption here is that each author  has unique writing styles that are relatively fixed and barely change with time. So we can use stylometric features to uniquely identify the writing style of the author~\cite{shalabi_kanaanbt}.\endgraf  

Along with the stylometric feature, we have used TF-IDF vectorizer and Word Embedding approach to identify depression from textual data. 

In this work, we are interested in identifying the depression of an author given a text. The Depression Identification(DI) problem has numerous applications in the emotional artificial intelligence field. We are interested in addressing the DI problem for social media Bangla text. To the best of our knowledge, only one paper has previously addressed this problem. 

Although depression identification has been widely studied in different languages, it is still understudied in the Bangla language. Bangla language is one of the most widely spoken and culturally rich languages. This language is the 7\textsuperscript{th} most spoken language~\cite{spoken_language} of the world and the native language of Bangladesh. However, this is not the only reason to study DI problems in the Bangla language. The problems associated with the Bangla language and the
relatively under-developed field of Bangla Natural
Language Processing (NLP) makes it more challenging to study
such problems for Bangla.

In this work, we will follow the following steps to study the DI problem. We will manually create a dataset from Facebook containing depressive posts and non-depressed posts. We will then pre-process the data and extract features from the data in 3 ways. One is called the Bag-of-words technique. Another one is computing stylometric features to capture the writing style of the author. The third one is using a word embedding approach to convert text into a feature vector.  

In recent years, deep learning-based recurrent neural models are used to  automate depression information extraction due to their performance in building models. These models do not require to be provided with pre-defined handpicked features. Instead, they can learn useful features from the data by themselves~\cite{Bsir_zrigui}.\endgraf  

In this work, we have used deep learning recurrent models to automate depression detection from Facebook textual data. Specifically, we have used LSTM and GRU models from deep learning recurrent neural network models. From the traditional model, we have used SVM and NB models. Then the performance of the deep learning model is compared to traditional machine learning models.

The rest of this paper is organized as follows: Section II overviews the related works of the depression identification problem. In section III, we have proposed our detailed solution. Section IV presents the experimental results. Finally, in section V, we conclude our findings  with a discussion of the obtained observations and the future directions of this work.

\section{Related Works}

In ~\cite{wang2013depression}, the authors conducted a depression analysis in Chinese. In their endeavor, Psychological and Machine Learning knowledge were combined. The authors opted for Psychologists who assisted 90 depressive and 90 non-depressed Micro-blog users in collecting a total of 6013 micro-blogs. Their model's precision was 80%. 

Abdul et al. ~\cite{abdul_hasib_arif} proposed using a Long Short-Term Memory Recurrent Neural Network (LSTM-RNN) to analyze Bangla social media posts for depression. They gathered Bangla tweets from Twitter in order to compile the dataset required for this endeavor. This dataset contained 1968 tweets, which was insufficient for a deep learning model to perform adequately. In order to improve the performance of the model with this short dataset, the data were stratified so that one depressive text was followed by one non-depressive text. They experimented with dividing the dataset into 80 percent for training, 10 percent for validation, and 10 percent for model testing. They optimized four hyperparameters of the trained model (LSTM size, batch size, number of epochs, and number of layers) for maximum classification accuracy. The evaluation of the model revealed an accuracy of 86.3\%.  

Authors performed the Gated Recurrent Neural Network algorithm on the same dataset in another work~\cite{uddin2019depression} to predict depressive Bangla text. As in previous work, They worked with the hyper-parameters of the GRU model and achieved approximately 75\% classification for this task. 

Billah et al.~\cite{billah2019depression} collected depressed and non-depressed posts from Facebook manually and applied SGD classifier, Multinomial Naïve Bayes, Logistic Regression, and Linear SVC to detect social media post whether it was depressive or not. During the treatment of patients,  Psychologists prefer some linguistic features that may help to detect depression. For example, depressed people usually use words like “me’, “I”, “myself” etc. which actually represent their self-centered thinking focusing on themselves rather than other people. The authors collected these types of posts to enrich their dataset which consisted of 1000 texts having depressive and not depressive texts. They applied several pre-processing steps to clean the data like punctuation removal, normalization, tokenization, etc. They applied Unigram, Bigram, and Emoticon features in their dataset. They had achieved the highest 77.9\% classification accuracy for the SGDC classifier.

Hassan et al. ~\cite{hassan_Jamil_2017sentiment} developed an automated system to detect depression levels of people from social media posts.  They removed the stop words and applied N-grams, POS tagging, and Negation feature extraction techniques to transform the text into a word vector. Finally, SVM, Naïve Bayes, and Maximum Entropy are applied to the dataset to classify depressive tweets where SVM showed the highest 91\% accuracy.

\section{Methodology}
\begin{figure}[htp]
    \centering
    \includegraphics[width=\columnwidth]{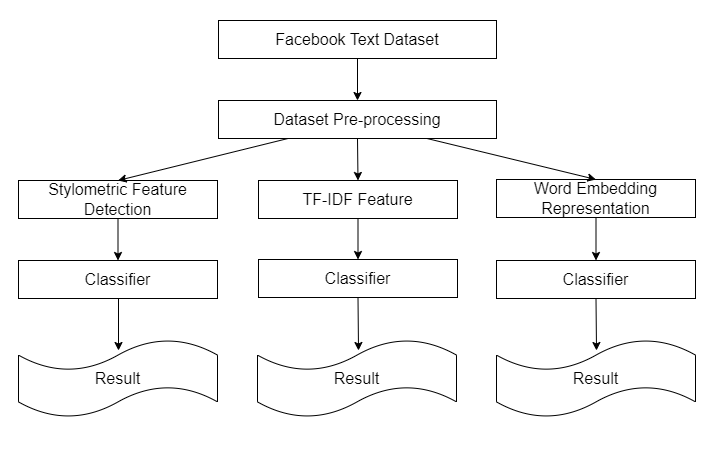}
    \caption{Steps followed in our solution}
    \label{fig:3.methodology}
\end{figure}

This section presents a detailed overview of three feature extraction techniques. One is the Bag-Of-Words feature and another is the Stylometric feature approach and the third one is the word embedding approach. We first label the dataset and then compute the feature after pre-processing the data. The feature is then fed into the machine learning model. Then we provide the architecture of the model.    Fig~\ref{fig:3.methodology} presents an overview of our proposed solution.  

\subsection{Dataset Creation}\label{AA}

Like any other TC problem, the first step in the depression detection problem is to build a large dataset. 
We have collected 981 texts from Facebook manually where the number of depressed class labels is 592 and the number of non-depressed class labels is 391 which is listed in Table~\ref{facebook_group_name_table}.\endgraf

\begin{table}[htbp]  
\caption{Data Distributions}
\begin{center}
\begin{tabular}{|p{0.2\textwidth}|p{0.2\textwidth}| }
\hline
 Class Label & Number of posts
 \\ 
 \hline
  Depressed Post & 
391
 \\ 
 \hline
Non-depressed Post &
592
 \\ 
 \hline
\end{tabular}
\label{facebook_group_name_table}
\end{center}
\end{table}

 We have taken the following facts in our concern during data collection: i) collecting the status of people who have committed suicide, ii)  Collecting the status of people who usually share depressive posts on social media, and iii) Collecting the status of people who share pleasant posts. For the labeling of data, we have taken 2 people's opinions to correctly determine the correct class label of data.

\subsection{Pre-processing}
The post obtained from Facebook is noisy and often contains a lot of unnecessary information. Here URL, image, tags etc. are present in the text. So we apply various pre-processing steps and have filtered out all characters except Bangla characters. Then we tokenize our texts and remove stop words from the text. We collect Bangla stop words from Github repository~\cite{git_stop_word_bangla} as mentioned in research work~\cite{tripto_ali}. 

Elongated words often contain some context in identifying emotion from the text. We express our feelings in elongated words. For example, "Greaaat news!!!" has more feelings than "Great news!!!". To maintain the context of the text, we do not apply lemmatization. 
\subsection{Word Vector Formation}
We use traditional machine learning models in our proposed solution. To use these models, we need to convert our text to a word vector. Conversion to word vector from text is done using TF-IDF features and Sentiment/Emotion Features approaches.    
\begin{enumerate}[wide, labelwidth=!, labelindent=0pt]

\item   
Stylometric Features Approach: \label{stylometric_feature_approach} \endgraf
   \hspace{10pt} Stylomtric features are features that capture the writing style of both depressed and non-depressed authors. We have already computed a large set of stylometric features based on existing works of~\cite{corney_anderson,Otoom_Amer}. These features are categorized into four types: lexical features, structural features, syntactic features and content-specific features. These four categories of stylometric features have been widely used in research work of ~\cite{abbasi_chen2,abbasi_chen}. Table~\ref{stylometric-features-table} lists all the 141 features we computed as stylometric features. \endgraf

 \hspace{10pt}Lexical features are the most common set of stylometric features that is intended for stylistics and text readability analysis. These features also signifies language assessment, first and second language acquisition. Lexical features consist of word-based and character-based features. These features are basically concerned with usage frequency of individual letters, vocabulary richness, entropy measure, consecutive occurrence of words etc. \endgraf
 
 \hspace{10pt}Syntactic features are primarily intended for identifying writing formation patterns such as the usage of punctuation marks. These features include total number of comma, colon, question marks and exclamation
marks etc.\endgraf

\hspace{10pt}Structure based features focuses on the way of organization of the layout of a text by an author. Organization of articles represents different habitual facts of an author such as paragraph length and use of greetings. As online texts have less content information but richer stylistic information, so these habits are seen to be more prominent in these texts in bearing strong authorial evidence of personal writing styles. We have computed 8 structure-related features as shown in Table ~\ref{stylometric-features-table}. \endgraf

\hspace{10pt}Finally, the content-specific features represents domain-specific terms. From the study of ~\cite{martindale_mcKenzie}, it has been shown that these features are important in the author's writing pattern formation. For these features, we have first got the feature words as suggested for the Arabic language in ~\cite{shalabi_kanaanbt}. Then we prepared the Bangla feature words by translating these Arabic words using Google Translator service API. We have translated Arabic words into 5 categories: Economy, Policy, Social, Sport, and Negative. % We have put some examples of this arabic words along with translated Bangla words in Table~\ref{arabic-translated-bangla}. 
This translation resulted in many duplicates, flaws, and inconsistencies in the translated lexicons. We have cleared all of these issues by manually inspecting the lexicons. In this way, we have prepared 5 content-specific features for the Bangla language listed in Table ~\ref{stylometric-features-table}. \endgraf

\hspace{10pt}For each text of the user, the feature extractor produced a 141 dimension vector to represent the values of the 141 features. As these feature sets contain information on the writing style of a user measured by various methods, the feature values we computed could range from 0 to any positive value. \endgraf

% For example, the first five feature values extracted from a text written by a male author could be 269 0.739 0.148 0.108 0.0 which represents the total number of Characters(C), the ratio of Bangla letters(total number of
% letters divided by C), the ratio of Bangla digits(total number of digits divided by C), the ratio of white space character(total number of white space character divided by C), the ratio of tab space character(total number of tab space character divided by C), respectively. 

\hspace{10pt}As we want to ensure all features are treated equally in the classification process, we have normalized the features using the max-min normalization method to ensure all feature values are between 0 and 1:

$$ x^{\prime}_{ij}  = \frac{x_{ij}-min(x_j)}{max(x_j)-min(x_j)} $$

where $x_{ij}$ is the \textit{j}th feature in the \textit{i}th example, min($x_j$) and max($x_j$) are the minimum and maximum feature values of the jth feature separately.

\begin{table*}[hp]
\caption{Stylometric Features}
\label{stylometric-features-table}
    \centering
    \begin{tabularx}{\textwidth}{| X | X | X |}
        \hline
       
        \multicolumn{3}{|c|}{Character Based Lexical Features} \\
        \hline
        Feature Name & Feature Count & Feature Description \\ 
        \hline
        Total characters $ \implies$ C   & $F_1$ & Alphabet, digits, special characters\\
        \hline
        Total number of letters/C & $F_2$ &  {\bng A} -  {\bng NN}  \\\hline
        Total number of digits/C & $F_3$ & {\bng 0}-{\bng 9} \\\hline
        Total number of white-space characters/C & $F_4$ & white sapce  \\\hline
        Total number of tab space characters/C & $F_5$ & tab space \\\hline
        Total number of elongation characters/C & $F_6$ & total word that pronounced as long \\\hline
        Total number of dependent vowel characters/C & $ F_7  \cdots F_{17} $ & {\bng a} , {\bng i} , {\bng ii} , {\bng u} ,  {\bng uu} ,{\bng rR} , {\bng e} ,  {\bng oi} ,  {\bng ou}  \\ \hline
        Total number of special characters/C & $ F_{18}  \cdots F_{41} $ &  
         \textless\textless, \textgreater\textgreater, \$, \%, \&, *, (, ), $>$, $<$, \{, \}, [, ], \_, +, -, =, \^ , \/,$ \backslash\backslash $, \textasciitilde, |          
        \\ \hline
        Total number of individual characters/C & $ F_{42}  \cdots F_{91} $ & {\bng A} -  {\bng NN}  \\
        \hline
        \multicolumn{3}{|c|}{Word Based Lexical Features} \\
        \hline
        Feature Name & Feature Count & Feature Description \\ 
        \hline
        Total number of words & $F_{92}$  & Total number of all words in the text  \\ \hline 
        Average word length & $F_{93}$ & The average number of characters per word \\ \hline
        Total Unique words & $F_{94}$ & Words that appear once in a text  \\ \hline
        Hapex legomena & $F_{95}$ & Words that occurs only one \\ \hline
        Hapax Dislegomena & $F_{96}$ & Words that occurs only twice \\ \hline
        Total no. of Short words & $F_{97}$ & 1-3 character words \\ \hline
        Total no. of Long words & $F_{98}$ & Words longer than 6 characters  \\ \hline
        Word length frequency distribution & $F_{99} \cdots F_{114} $ & word length from 0 to 15  \\ \hline
        Yules K measure & $F_{115}$ & Measure of vocabulary richness  \\ \hline
        Total Digit word & $F_{116}$ & words that have digits \\ \hline
        Total Consecutive occurrence & $F_{117}$ & no of words with 3 consecutive duplicate characters \\ \hline 
        
         \multicolumn{3}{|c|}{Syntactic Features} \\
        \hline
        Feature Name & Feature Count & Feature Description \\ 
        \hline
        Total single quotes/C   & $F_{118}$ & ' \\
        Total comma/C & $F_{119}$ & ,  \\
        Total periods/C & $F_{120}$ & . \\
        Total colons/C & $F_{121}$ & :  \\
        Total semi colons/C & $F_{122}$ & ; \\
        Total question marks/C & $F_{123}$ & ? \\
        Total exclamation marks/C & $F_{124}$ & ! \\
        Total double quotes/C & $F_{125}$ &  "  \\ 
        \hline

\multicolumn{3}{|c|}{Structural Features} \\
\hline
        Feature Name & Feature Count & Feature Description \\ 
        \hline
        Total number of lines & $F_{126}$ & line splitted by {\bng .} \\
        Total number of paragraphs & $F_{127}$ & in the case of pressing enter \\
        Total number of sentences & $F_{128}$ &  \\
        Total number of blank lines & $F_{129}$ & in the case of pressing enter \\
        Total number of non blank lines & $F_{130}$ & \\
        Total length of non blank lines & $F_{131}$ & \\
        Average length of non blank lines & $F_{132}$ & \\
        Average number of words per sentence & $F_{133}$ & \\
        Average number of words per paragraph & $F_{134}$ & \\
        Average number of character per sentence & $F_{135}$ & \\
        Average number of  character per paragraph & $F_{136}$ &  \\
        \hline  
        
        \multicolumn{3}{|c|}{Content Specific Features} \\
\hline
Feature Name & Feature Count & Feature Examples \\ 
\hline
        Total economic phrases & $F_{137}$ & {\bng rajY tHibl } (State funds), {\bng ekam/pain } (Company), {\bng ibineyag }(Investment) \\
         \hline 
        Total policy phrases & $F_{138}$ &  {\bng raSh/TRduut}(Ambassador),{\bng ibshrRNG/khl}(Chaotic), {\bng Jt/n}(Care) \\
         \hline 
        Total Social phrases & $F_{139}$ &  {\bng narii Aidhkar} (Women's rights), {\bng bNNGsh}(race), {\bng Jub gRup}(youth group) \\
         \hline 
        Total Sports phrases & $F_{140}$ & {\bng ekhla} (Sports), {\bng fuTbl}(Football), {\bng pdk}(Medal)\\
         \hline 
        Total Negative phrases & $F_{141}$ & {\bng pirtYag kora} (abandon), {\bng ghrRNa}(hatred), {\bng polayon}(escape)\\
        \hline 
    \end{tabularx}
\end{table*} 

 % Total economic phrases & $F_{137}$ & রাজ্য তহবিল (State funds), কোম্পানি (Company), বিনিয়োগ(Investment) \\
 %         \hline 
 %        Total policy phrases & $F_{138}$ & রাষ্ট্রদূত(Ambassador), বিশৃঙ্খল(Chaotic), যত্ন(Care) \\
 %         \hline 
 %        Total Social phrases & $F_{139}$ & নারী অধিকার(Women's rights), বংশ(race), যুব গ্রুপ(youth group) \\
 %         \hline 
 %        Total Sports phrases & $F_{140}$ & খেলা(Sports), ফুটবল(Football), পদক(Medal)\\
 %         \hline 
 %        Total Negative phrases & $F_{141}$ & পরিত্যাগ করা(abandon), ঘৃণা(hatred), পলায়ন(escape)\\
 %        \hline 

\item   
\newcolumntype{s}{X}
\newcolumntype{b}{>{\hsize=1.5\hsize}X}

\begin{table*}[t]
 \caption{Example Sentence for TF-IDF Vectorizer 
}
\label{tf-idf-text-example}
\begin{tabularx}{\textwidth}{sb}
\toprule
Text No. & Example\\
\midrule
Text1 &  I love natural language processing.\\ 
Text2 & I like data processing. \\ 
Text3  & I like text processing. \\ 
\bottomrule
\end{tabularx}
\end{table*}

\begin{table*}[t]
 \caption{Example Sentence for TF Vectorizer 
}
\label{tf-vectorizer-example}
\begin{tabularx}{\textwidth}{|s|s|s|s|s|s|s|s|s|}
\toprule
Text No. & I & love & like & natural & language & data  & text  & processing\\
\midrule
Text1 &  1 & 1 & 0 & 1 & 1 & 0 & 0 & 1 \\ 
\midrule
Text2 & 1 & 0 & 1 & 0 & 0 & 1 & 0 & 1  \\
\midrule
Text3  & 1 & 0 & 1 & 0 & 0 & 0 & 1 & 1 \\ 
\bottomrule
\end{tabularx}
\end{table*}

\begin{table*}[t]
 \caption{Example Sentence for IDF Vectorizer 
}
\label{idf-vectorizer-example}
\begin{tabularx}{\textwidth}{|s|s|s|s|s|s|s|s|s|}
\toprule
No. & I & love & like & natural & language & data & text & processing\\
\midrule
Word &  0 & 0.477 & 0.18 & 0.477 & 0.477 & 0.477 & 0.477 & 0 \\ 
\bottomrule
\end{tabularx}
\end{table*}

\begin{table*}[t]
 \caption{Example Vector for TF-IDF Vectorizer 
}
\label{tf-idf-vectorizer-example}
\begin{tabularx}{\textwidth}{|s|s|s|s|s|s|s|s|s|}
\toprule
Text No. & I & love & like & natural & language & data & text & processing\\
\midrule
Text1 &  0 & 0.477 & 0 & 0.477 & 0.477 & 0 & 0 & 0  \\ 
\midrule
Text2 & 0 & 0 & 0.18 & 0 & 0 & 0.477 & 0 & 0  \\
\midrule
Text3  & 0 & 0 & 0.18 & 0 & 0 & 0 & 0.477 & 0 \\ 
\bottomrule
\end{tabularx}
\end{table*}

TF-IDF Count Vectorizer Approach: \endgraf

\hspace{10pt} TF-IDF stands for Term Frequency Inverse Document Frequency. TF-IDF converts the text of the user into a meaningful number of vectors to fit in the machine learning model. In any document, the term frequency is the number of occurrences of a term. On the other hand, document frequency represents the number of documents containing of that term.  Term frequency indicates the importance of a specific term in a document. Document frequency indicates how common the term is in~\cite{tf_idf_link}. TF-IDF gives higher weights to terms appearing frequently
in the given document and rarely in other documents.

We have implemented the TF-IDF count vectorizer from the scikit-learn library. Since, TF-IDF generates higher dimensional feature vectors, we need dimensional reduction techniques. We can do these in several ways. The first one is to reduce the number of extracted terms. Another one is to use stemming. In our work, we have used the reduction of extracted terms approach. We set the number of extracted terms as 1000.    

For Example, let us take an example. Suppose, we have 3 texts which have to be converted to vectors using TF-IDF count vectorizer as shown in the table Table~\ref{tf-idf-text-example}. 

For converting these texts into vectors, we have to first identify unique words and count how many times these words occur in each text. This is shown in Table~\ref{tf-vectorizer-example}

Then we have to compute inverse document frequency (IDF) using the following formula. 

$$ idf_i = log\frac{n}{df_i} $$

where $df_i$ represents how many documents contain the term i and n is the total number of documents. We have calculated the inverse document frequency for each work and shown it to Table~\ref{idf-vectorizer-example}

Now we will multiply the TF matrix with IDF score to get the vectorized form of each text which is shown in Table~\ref{tf-idf-vectorizer-example}. We have converted all texts into vectors. These vectors can be fed into any machine-learning algorithm.

\item   

Word Embedding Representation Approaches: \endgraf
   \hspace{10pt}Traditionally, the bag-of-words (BOW) model is used to transform the text into feature vectors in text classification.  In this model, a text is represented as the bag (multiset) of its words, disregarding grammar and even word order but keeping multiplicity. Authors in~\cite{shalabi_kanaanbt} followed the BOW model with a set of hand-crafted rules to prepare the feature set.  \endgraf
   
   However, motivated by the recent
the success of deep learning models in text classification, we have used word embeddings as features instead of applying the BOW model. Word embeddings are a distributed representation of text  that allows words with similar meanings to have a similar representation. In this approach, individual words are represented as real-valued vectors in a predefined vector space. Authors in ~\cite{mikolov_zweig} have shown that word embeddings can capture syntactic and semantic regularities employing the vector offsets between word pairs sharing a particular relation. Below, we explain how Word2vec is used to generate word embeddings. 

\begin{enumerate}[wide, labelwidth=!, labelindent=0pt]
 \item Word2vec:
 
 Word2vec, an efficient algorithm proposed by Google~\cite{mikolov_kai},  can  learn a standalone word embedding from a text corpus efficiently maintaining the contextual meaning of words. Word2vec has two model architectures to produce an embedding representation of words. One is Continuous Bag of Words(CBOW) and another is Skip Gram (SG). We have implemented both of these model architectures as mentioned in ~\cite{mikolov_kai}. \endgraf 
\endgraf
 CBOW Model takes the context of each word as the input and tries to predict the word corresponding to the context. SG predicts the surrounding window of context words based on the current single word. The word vector prediction is not influenced by the order of context words. Figure~\ref{fig:cbow_skip_gram_model_architecture} shows the architecture of the Skip-Grams and Continuous Bag of Words model. Below we have described in detail how CBOW and Skip-Gram Model is used to generate word vector from the text. \endgraf  

\begin{enumerate}
  \item CBOW Model: CBOW model predicts the current single word from a specified window of surrounding context words. To generate word embedding, we first select a vocabulary of size V from our dataset. Then, we create a one-hot encoded vector for each of these words in \textit{V}. For a window size \textit{C}, the sliding window method generates features for every single word in each training sample to train the CBOW model using the generated features. 
  
  Then, for each training sample, we feed forward one hot encoded vector of input words to a Neural Network consisting of one hidden layer with N nodes. The output layer has V nodes with a softmax activation function. The output word vector is then compared against the actual word vector. The weight of each layer is updated according to the error and backpropagation happens. Figure~\ref{fig:cbow_architecture} shows the architecture of the Continuous Bag of Words model. 
  
  \textbf{Example 1.} "Have a great day"- take the word “great” as an example. Assuming windows size as one, training features are- ([Have, great], a), ([a, day], great). CBOW model learns to predict 'day' in the context given 'a' and 'great' as the input.
  
  Our vocabulary is \{Have, a, great, day\}. The encoded one hot vectors are: Have=\{1, 0, 0, 0\}, a=\{0, 1, 0, 0\}, great=\{0, 0, 1, 0\}, day=\{0, 0, 0, 1\}. 
  
  For this training feature, ([a, day], great), the CBOW model first feeds forward the one hot encoded vector of 'a' and day. \textit{$W_{VN}$} is multiplied with these vectors. Then after doing the average, we get the hidden layers vector. \textit{$W`_{NV}$} is the weight matrix that maps the hidden layer outputs to the final output layer. This output layer provides the word vector of great. This word vector of great is then compared to the actual word vector of great and \textit{$W_{VN}$} and \textit{$W`_{NV}$}
vectors are updated.\endgraf   
 
   \item Skip-Gram model: Skip-Gram model is a neural network that can generate word embedding without any labeled data. This neural network does so by creating a “fake” task to train. We aren't interested in the input and output of this neural network rather than the goal is to learn the weights of the hidden layer that are actually the “word vectors” of the corresponding words. 
 
 The fake task for the Skip-gram model would be, given a word, we’ll try to predict its neighboring words. We’ll define a neighboring word by the window size — a hyper-parameter.
 
 Given a sentence: “The quick brown fox jumps over the lazy dog.” and a window size of 2, if the target word is brown, its neighboring words will be ( the, quick, fox, jumps). Our input and target word pair would be (brown, the), (brown, quick), (brown, fox), (brown, jumps). Within the sample window, proximity of the words to the source word does not play any role. So the, quick, fox, and jumps will be treated the same while training. Figure~\ref{fig:skip_gram_architecture} shows the architecture of Continuous Bag of Words model. 
 
 We will first select a vocabulary of size V from our dataset. Then we create a one-hot encoded vector of dimension 1xV for each of these words in V. So the dimensions of the input vector in Skip-Gram model will be 1xV. The single hidden layer will have dimension VxN, where N is the size of the word embedding and is a hyper-parameter. The output from the hidden layer would be of the dimension 1xN, which we will be fed into an softmax layer. There  will be C vector in the output layer. The dimension of each vector in output layer will be 1xV, where each value in the vector will be the probability score of the target word at that position.

For training samples corresponding to a source word, we will back propagate in one back pass. So for brown, we will complete the forward pass for all 4 target words ( the, quick, fox, jumps). We will then calculate the errors vectors[1xV dimension] corresponding to each target word. We will now have 4 1xV error vectors and will perform an element-wise sum to get a 1xV vector. The weights of the hidden layer will be updated based on this cumulative 1xV error vector. 
 
\end{enumerate}

Both Skip-gram and CBOW models can be used on a large dataset to learn word embedding in a short time like billions of
words in hours. This can be achieved in a low computational complexity. Practically, Skip-gram gives better word representations for small data-set. CBOW provides better performance and is more suitable
when the datasets is large~\cite{mikolov_kai}. Further details of the Skip-Grams and Continuous Bag of Words model can be found in ~\cite{svoboda_brychcín}

% \begin{figure*}[t!]
%       \centering
% \begin{subfigure}[t]{0.9\textwidth}
% \includegraphics[width=\textwidth]{Tex_Code/resource/image/skip_cbow.PNG} 
% \end{subfigure} 
% \caption{Architecture of Skip-Gram and Continuous Bag of Words Model }
%   \label{fig:3.c.2.skigram_cbows}
% \end{figure*} 

\item Word Embedding using Word2vec: \endgraf  
As per research study in~\cite{mikolov_kai}, Skip-gram and Continuous Bag of Words, two architectures of Word2vec model, can be used on any large corpus of text to learn the word embedding. As Word2vec model can capture a lot of information maintaining semantic, conceptual and contextual relation, we have learned the embedding vector of each word from user post from Facebook of our dataset using this CBOW and Skip-Gram model. \endgraf

For example, let us have two sentences in our dataset. [I have a book, I love to eat mango]. So first we split the sentence and generated a two-dimensional vector. The two-dimensional vector will be [ [I, have, a, book], [I, love, to, eat, mango]]. Then we pass this two-dimensional vector to the word2vec model. Skip-Gram and CBOW model generates the word vector from this two-dimensional dataset using  window-size  as 5. The size of the word vector is 300.   We have used the Gensim package to implement the  Word2vec model. This model returns a 300-dimensional vector for each of these words: I, have, a, book, love, to, eat, mango. We save these word embedding and later used these embeddings as the weight of the embedding layer. \endgraf
     
\item Classification:

For a labeled sentence $\braket{s,c}$ , we first encoding each word of sentence s using a unique number. For example, if 1 is assigned to "cat", 2 to "mat", and so on, then encoding of the sentence "The cat sat on the mat" as a dense vector will be like [5, 1, 4, 3, 5, 2]. \endgraf 
Then we multiply this encoding vector with the embeddings of words present in s to form the hidden representation of s. These sentence representations are used to train a linear classifier. Specifically, we use the softmax
function to compute the probability distribution over the classes in C. 

For example, Lets say our training sample consists of two sentences: "I want to see you again." and "Feeling better to see you again". If we encode these phrases by assigning each word a unique integer number, then our phrases could be rewritten as: [0, 1, 2, 3, 4, 5] , [6, 7, 2, 3, 4, 5]. We pass this training samples to embedding layer which can be defined as below: 

\inlinecode{Java}{Embedding(8, 2, input_length=6)}

The first argument (8) is the number of distinct words in the training set. The second argument (2) indicates the size of the embedding vectors. The input-length argumet determines the size of each input sequence.

The word embedding of each of these words present in our dataset is got from Word2vec model and is set as weight of
embedding layer. Embedding layer can be thought as the table used to map integers to embedding vectors. Lets say, currently embedding layer has embedding vectors defined in Table~\ref{embedding_layer_weight}: 

So according to these embeddings, our second training phrase will be represented as:

 [[4.1, 2.0], [2.1, 3.2], [1.0, 3.1], [0.3, 2.1], [2.2, 1.4], [0.7, 1.7]] 
 
 After passing through embedding layer, each of our training samples will be converted into 2-dimensional vector which is gone into LSTM model.

\begin{center}
\begin{table}[h!]
\centering
\begin{tabular}{||c c c ||} 
 \hline
 word & index & Embedding\\ [0.5ex] 
 \hline\hline
 I & 0 &  [1.2, 3.1] \\ 
 \hline
 want & 1 & [0.1, 4.2]   \\
 \hline
 to & 2 & [1.0, 3.1] \\
 \hline
 see & 3 & [0.3, 2.1]\\
 \hline
 you & 4 & [2.2, 1.4] \\   
 \hline
 again & 5 & [0.7, 1.7] \\   
 \hline
 Feeling & 6 & [4.1, 2.0] \\   
 \hline
 better & 7 & [2.1, 3.2] \\ [1ex] 
 \hline
\end{tabular}
\caption{Sample weight of Embedding Layer}
\label{embedding_layer_weight}
\end{table}
\end{center}

The output from LSTM layer goes to flatten layer which converts the two dimensional vector into 1-dimensional vector. If the 2-dimensional vector of second training phrase is passed through flatten layer, then the layer will give output:  [4.1, 2.0, 2.1, 3.2, 1.0, 3.1, 0.3, 2.1, 2.2, 1.4, 0.7, 1.7]. The output from flatten layer is passed through dense layer. 

Dense layer the regular deeply connected neural network layer which computes output = activation(dot(input, kernel) + bias). Dense layer is providing two valued 1 dimensional vector. First value depicts the probability of being male and second value depicts the probability of being female. Softmax is used as activation function in dense layer whose computation is done using the following formula: 

$$\text{Softmax}(x_{i}) = \frac{\exp(x_i)}{\sum_j \exp(x_j)}$$

\end{enumerate}

\begin{figure*}[ht!]
       \centering
\begin{subfigure}[t]{0.4\textwidth}
\includegraphics[width=\textwidth]{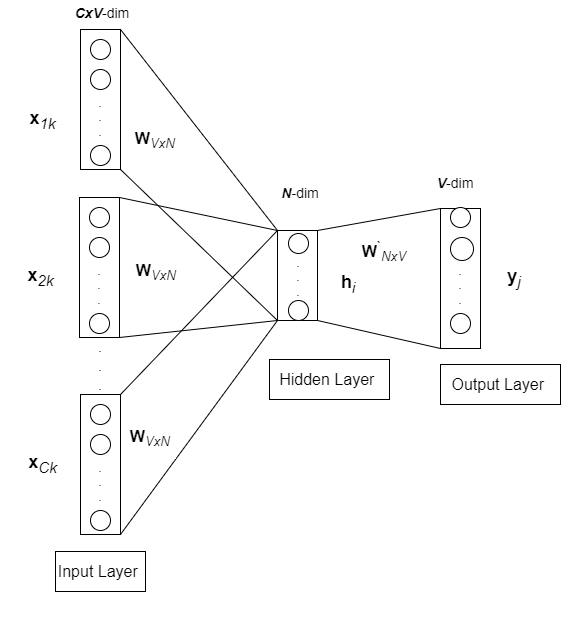}

\caption{CBOW model architecture} \label{fig:cbow_architecture}
\end{subfigure}
\begin{subfigure}[t]{0.4\textwidth}
\includegraphics[width=\textwidth]{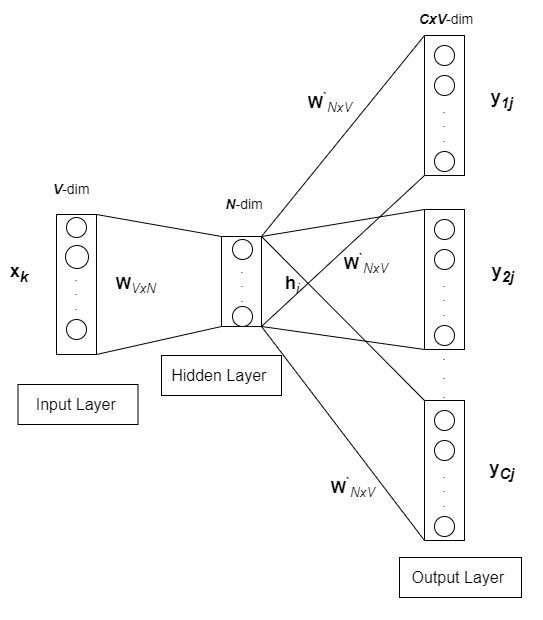}
\caption{Skip-Gram model architecture}
\label{fig:skip_gram_architecture}
\end{subfigure}
\caption{Architecture for CBOW and Ski-Gram Model}
  \label{fig:cbow_skip_gram_model_architecture}
\end{figure*}
    
% A word cloud representation of % our vocabulary is given in % Fig.~\ref{fig:wordcloud}.

% To get word vectors into embedding form, we have followed the following steps:

% \begin{enumerate}
%     \item First we select a vocabulary of size D from our text corpus. For example, for the following two sentences, let the value of $MAX\_WORDS$ is 5. Dataset is  [`I love to buy books',`I love to sell books']
    
%     \item The words present in the dataset are 'I', 'love', 'to', 'buy', 'books', 'sell'. As vocabulary size D is 5, so we have taken these 5 words, 'I', 'love', 'to', 'buy', 'books' based on their frequency. 
    
%     \item For every sentence in our Dataset, we convert the text into word vector of size D based on their index. As a result, our word vectors will be [[1,2,3,4,5],[1,2,3,5]]
    
%     \item These word vectors are padded to a vector of fixed length. So our word vectors will be [[1,2,3,4,5],[1,2,3,0,5]]. Then these vectors are passed to an embedding layer. 
    
%     \item We have prepared 2D list of words from our dataset and passed this words list to word2vec algorithm. Word2vec algorithm generates real valued vectors for each word which are set as weight matrix of above embedding layer.
    
%     \item We get word vector from this embedding layer that is close to each other having the similarity in context.
    
% \end{enumerate}

\end{enumerate}
\subsection{Model Architecture}
For two types of features, we have implemented deep learning models and traditional machine learning models to detect depression in the text. In the deep learning model, we implement LSTM and GRU models. In the traditional machine learning model, we implement SVM, and NB models. 

\begin{figure*}[ht!]
\centering
\begin{subfigure}[t]{0.8\textwidth}
\includegraphics[width=\textwidth]{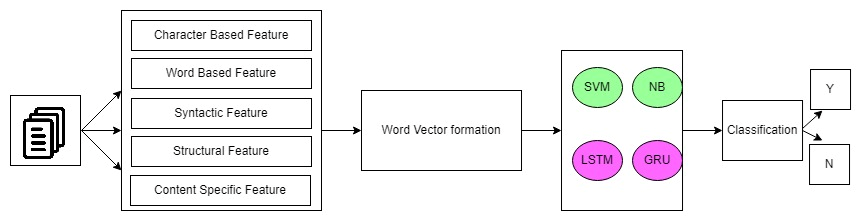}
\caption{Model Architecture for Stylometric Feature} \label{fig:model_architecture_for_stylometric_feature}
\end{subfigure}
\bigskip

\begin{subfigure}[t]{0.8\textwidth}
\includegraphics[width=\textwidth]{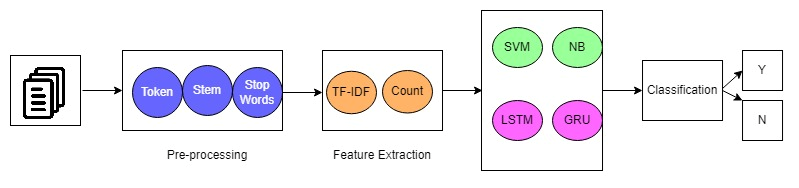}
\caption{Model Architecture for TF-IDF}
\label{fig:model_architecture_for_TF-IDF}
\end{subfigure}

\begin{subfigure}[t]{0.8\textwidth}
\includegraphics[width=\textwidth]{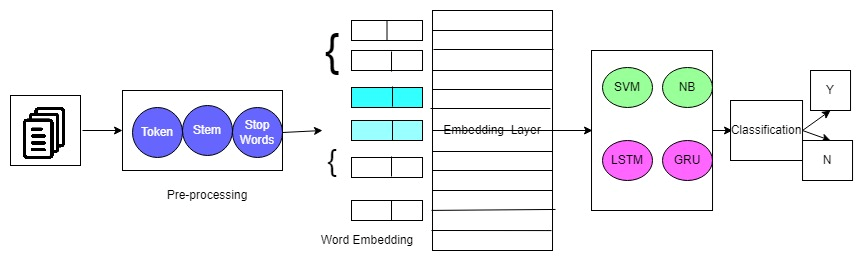}
\caption{Model Architecture for Word Embedding}
\label{fig:model_architecture_for_word_embedding}
\end{subfigure}
\caption{Model Architecture}
\vspace{-30pt}
\end{figure*}

\begin{enumerate}
  \item Model with Stylometric Features: \\
     We have prepared word vectors by Styometric features discussed in Section-IIIC. We pass these vectors to the LSTM layer having 300 nodes. The output of the LSTM layer is passed to a dense layer. Softmax~\cite{sharma_ochin} is used as an activation function. The optimizer is RMSprop and binary cross entropy ~\cite{shipra_saxena} is used as a loss function. The same process is repeated for Gated Recurrent Unit(GRU), SVM, and NB models. We have noted down each model's accuracy, F1-score.
     Fig. \ref{fig:model_architecture_for_stylometric_feature} shows the architecture of the LSTM, GRU, SVM, and NB model with the stylometric feature.

  \item Model with TF-IDF feature:\\ 
  We have implemented deep learning models like LSTM, GRU, and traditional machine learning algorithms such as Support Vector Machine(SVM) and Naive Bayes (NB) to detect depression from the text. We have used Term Frequency Inverse Document Frequency (TF-IDF) to generate a feature set from the text. 2 grams is used in the TF-IDF vectorizer. We have used the linear kernel function in the Support Vector Machine model. The function of kernel is to take data as input and transform it into the required form. For each model, we have noted down accuracy and F1-score. 

   \item Model with Word Embedding feature:\\ 
 For any sentence S with classification C, we have done the necessary prepossessing as discussed in Section-IIIC. Then these sentences are passed through a tokenizer which can produce a one-hot encoding vector of length 100. Only the top 1000 most frequent words are taken as vocabulary. The first 100 words are taken for sentences having more than 100 words. Shorter text is padded with zeros. After that, these vectors are fed into an embedding layer. The weights of the embedding layer are initialized with word2vec embedding weights. We initialize the embedding layer using trainable and non-trainable properties. Non-trainable property freezes the Embedding layer so that the pre-trained weights are not updated during the training. The output dimension of the embedding layer is 300 as it is the vector length of each word in the word2vec model. The sequence of 100 words is then passed through an LSTM layer. The output of the LSTM layer is passed to a dense layer which is used to detect depression in the text. Softmax~\cite{sharma_ochin} is used in the dense layer as an activation function. The optimizer is RMSprop and binary cross entropy~\cite{shipra_saxena} is used as a loss function. The same process is repeated for all the remaining deep-learning models. Figure~\ref{fig:model_architecture_for_word_embedding} shows the architecture of our models with the word embedding features.  
\end{enumerate}

\section{Experimental Evaluation}
In this section, we have evaluated the performance of our proposed methods for depression detection on Facebook data sets. We compare the performance of deep learning algorithms with traditional machine learning algorithms like Support Vector Machine(SVM) and Naive Bayes(NB). 

\subsection{Experimental Setup}
We have employed Python Keras framework with Tensorflow as a framework to implement deep learning models for training, tuning, and testing. We have also used Gensim package for word2vec model implementation. Scikit-learn package is used to implement traditional machine learning algorithms. Experimental evaluation was conducted on a machine with an Intel Core i7 processor with 1.8GHz clock speed and 8GB RAM. The machine has also an Nvidia GeForce MX150 with 2GB memory and therefore Tensorflow based experiments have fully utilized GPU instructions. Considerable speed can be achieved in Tensorflow based experiments by adding a GPU as shown in ~\cite{JMLR_speed_gpu}. 
\subsection{Performance Evaluation and Parameter Tuning}
We have studied the efficiency and scalability of our proposed methods by varying model architectures and feature set vectors. We have measured performance of recurrent neural network, LSTM model by generating word vector using stylometric feature method and using word2vec method. Moreover, we have evaluated the performance of different word2vec model by initializing embedding layer weights with them and consider both trainable and non-trainable weight matrix. Trainable weight matrix denotes that at the time of training, the embedding layer weights will be updated. Non-trainable weight matrix denotes  that the embedding layer weights will not be updated at the time of training.\endgraf 
 
\subsection{Result Analysis}

\begin{table*}[t]
 \caption{Performance measure for Stylometric Feature: 
}
\label{performance-table-stylometric-feature}
\begin{tabularx}{\textwidth}{@{}l*{2}{C}c@{}} 
\toprule
Model & Feature Vector & Accuracy & F1-score \\
\midrule
LSTM & Styoletric Feature & 76.42\% & 34.35\% \\ 
GRU & Styoletric Feature & 74.58\% & 34.36\%  \\ 
SVM & Styoletric Feature & 72.95\% & 31.91\% \\ 
NB & Styoletric Feature & 75.96\% & 37.30\% \\ 
\bottomrule
\end{tabularx} 
\end{table*}

\begin{table*}[t]
 \caption{Performance measure for Word Embedding Feature: 
}
\label{performance-table-embedding-feature}
\begin{tabularx}{\textwidth}{@{}l*{2}{C}c@{}} 
\toprule
Model & Feature Vector & Accuracy & F1-score \\
\midrule
LSTM & Word Embedding Feature & 49.33\% & 24.45\% \\ 
GRU & Word Embedding Feature & 41.93\% & 23.23\%  \\ 
SVM & Word Embedding Feature & 55.32\% & 22.22\% \\ 
NB & Word Embedding Feature & 62.94\% & 23.16\% \\ 
\bottomrule
\end{tabularx} 
\end{table*}

\begin{table*}[t]
 \caption{Performance measure for TF-IDF Feature: 
}
\label{performance-table-tfidf-feature}
\begin{tabularx}{\textwidth}{@{}l*{2}{C}c@{}} 
\toprule
Model & Feature Vector & Accuracy & F1-score \\
\midrule
LSTM & TF-IDF Feature & 60.91\% & 29.41\% \\ 
GRU & TF-IDF Feature & 57.86\% & 33.6\%  \\ 
SVM & TF-IDF Feature & 65.92\% & 41.74\% \\ 
NB & TF-IDF Feature & 65.99\% & 43.7\% \\ 
\bottomrule
\end{tabularx} 
\end{table*}

\begin{table*}
\setcellgapes{3pt}
\makegapedcells
\caption{Comparison of Proposed Method with Previous Works}
\label{table:comparison_of_proposed_method_with_previous_work}
\begin{tabularx}{\textwidth}{ | *{5}{C|} }
  \hline
  \textbf{Language} &  \textbf{Proposed by} &  \textbf{Methodology} &  \textbf{Feature Extraction} & \textbf{Accuracy} \\ 
  \hline 
  Bangla & ~\cite{abdul_hasib_arif}  & LSTM  model &
  None
  & 86.3\%   \\
    \hline
   English & ~\cite{Rafiqul_kabir_Raihan} & KNN, SVM, DT
& LIWC
& 73\%  \\
    \hline
    Bangla & ~\cite{billah2019depression}  & SGD Classifier, Logistic Regression
 & Unigram and Emoticons with TF-IDF values
 & 74.57\%  \\
    \hline
    Bangla & ~\cite{uddin2019depression} &  GRU  & None & 75.7\% \\
    \hline
     English & ~\cite{hassan2017sentiment}  & SVM, Naïve Bayes, ME & N-gram, POS, Negation Checker & 91\%  \\
    \hline
Bangla &  Proposed Method & LSTM, GRU, SVM, NB & Stylometric Feature, TF-IDF Feature, Word Embedding Feature &  76.42\%\\
    \hline
  
 \end{tabularx}
\end{table*}

\begin{enumerate}
\item Performance of stylometric feature: 

We present the performance of stylometric feature along with traditional machine learning and deep learning algorithms in Table~\ref{performance-table-stylometric-feature}. From Table~\ref{performance-table-stylometric-feature}, we can see that deep learning algorithm like LSTM  performs better than traditional machine learning algorithms such as SVM \& NB. We have used stylometric features that can capture the writing style of different authors.  NB outperforms the SVM model because the NB model with stylometric features can predict depression based on probabilistic distribution.  Again, LSTM having feedback connections can process the entire sequence of data. Thus LSTM with stylometric features outperforms traditional machine learning algorithms like SVM and NB.

\item Performance of word embedding feature:

We have experimented using word embedding as a feature with both deep learning and traditional machine learning algorithms. The performance comparison is  
noted down in Table~\ref{performance-table-embedding-feature}.  The performance of NB \& SVM models with word embedding features outperforms LSTM and GRU models. NB models can predict class labels based on probabilistic distribution. Also, we have used Word2vec~\cite{mikolov_kai} which can generate word vectors based on the context. Thus the performance of NB model in combination with the word embedding feature is better than deep learning models.    
\item Performance of TF-IDF feature: 

The performance of the TF-IDF feature vector in accordance with deep learning algorithms like LSTM, GRU and traditional machine learning algorithms such as SVM, and NB is presented in Table~\ref{performance-table-tfidf-feature}. Here, the NB model outperforms SVM, GRU, and LSTM models. TF-IDF feature representation, a bag of words(BOW) feature,  is used to assign weight to a word based on the number of times it appears in the text. As the NB model predicts the output class based on the probability of the given vector, so NM model with the TF-IDF feature outperforms another model. The highest accuracy is 75.96\% got by the NB model with the TF-IDF feature. 

  \item Performance among different models:    

  We present the performance of different machine learning model associated with the feature vector in Figure~\ref{fig:performance_compare_among_different_models}. This figure is a summarising table of Table~\ref{performance-table-stylometric-feature},~\ref{performance-table-embedding-feature}, and~\ref{performance-table-tfidf-feature}.If we take best performance of machine learning algorithm in terms of feature, then we will get Figure~\ref{fig:performance_compare_among_different_models}. From Figure~\ref{fig:performance_compare_among_different_models}, we can see that deep learning model LSTM model outperforms traditional machine learning models such as SVM \& NB. LSTM with the stylometric feature is slightly better than NB with the stylometric feature. This is expected as LSTM can save previous state. NB is slightly better than SVM in terms of accuracy since the NB can predict class labels based on the simpler computation of text frequency to compute the posterior probability of each class. On the other hand, SVM works based on the computation of hyperplane equations which separates data into classes perfectly~\cite{Kusumawati_2019}. The highest accuracy and F1-score got by our proposed methods are 76.42\% and 34.35\% accordingly.

% \begin{figure*}[ht!]
% \centering
\begin{figure}[t]
\centering
\includegraphics[width=0.5\textwidth]{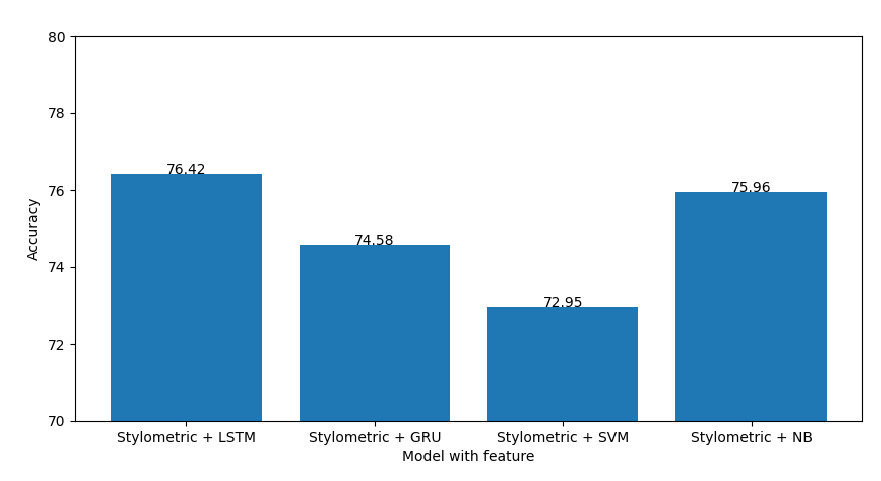}
\caption{Performance compare among different models} \label{fig:performance_compare_among_different_models}
\end{figure}
% \end{figure*}

\item Training and Validation Accuracy for LSTM:  
As the stylometric feature with the LSTM model outperforms all other models, we have drawn a plot on training accuracy vs validation accuracy for the LSTM model. From the figure~\ref{fig:training_and_validation_accuracy}, we can see, with an increase in epochs, both training and validation accuracy increase. But validation accuracy is higher than training accuracy in this case. The reason is that during random splitting, the examples that are set in the validation set are easier to guess than in the training set.        

\begin{figure}[h!]
\centering
\includegraphics[width=0.5\textwidth]{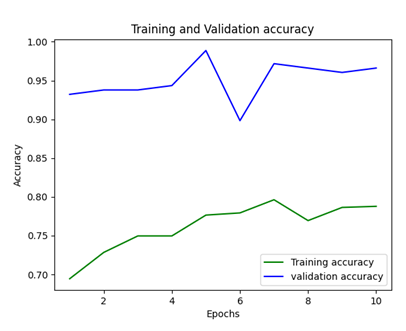}
\caption{Training and Validation Accuracy for LSTM} \label{fig:training_and_validation_accuracy}
\end{figure}

\item Comparison of Proposed Method with Previous Works: 

In Table ~\ref{table:comparison_of_proposed_method_with_previous_work}, we have shown our proposed methodology and previous works done on depressed port detection from social media text both in English and Bangla language. As the number of previous works done on depressed post detection from social media text in the Bangla language is very low, among ~\cite{abdul_hasib_arif}, ~\cite{billah2019depression},  ~\cite{uddin2019depression} and our proposed methodology which works are done for Bangla language, our work is showing comparatively better performance.

% \item Performance of deep learning model vs traditional machine learning model:

% The comparison of performance between traditional machine learning(SVM, NB) model and the deep learning(LSTM) model is presented in Table~\ref{performance-table-different-model}. Foe feature representation, we have compared both traditional representations such as TF/IDF which is a variant of BOW and embedding representations such as Word2vec. Even though the drawback of BOW approach is the the curse of dimensionality, the BOW approach has still the power to solve solve our gender problem. TF/IDF vectorizer performs better than Context based Word2vec vectorizer approach. From Table~\ref{performance-table-different-model}, we can see that traditional ML models show better performance than neural network model. Widely used algorithms such as SVM and NB have proved to be efficient and successful for author gender identifications. SVM algorithm can segregate text according to gender by creating decision boundary. 
% We have used linear decision boundary. Naive Bayes classifiers categorize text data based on strong independence assumptions between the features. On the other hand, neural network model such as LSTM categorize data based on weights which seems to be under-performed in our study. This result is consistent with research work ~\cite{sboev_rybka,T_YILDIZ}

\end{enumerate}

\section{Conclusion \& Future Work}

Depression detection from social media text is one of the research work done in various languages. In the Bangla language, there is no standard dataset for depression detection. We have made our dataset public so that future work can be done using this dataset. In this work, we have applied deep learning models and traditional machine learning models to detect depression from Bangla text. Three types of features such as stylometric feature, TF-IDF feature, and Word embedding feature are used. Among them, LSTM with stylometric features shows the highest accuracy which is 76.42\%. This work will help us to detect depressed people within our network before they go under a malignant stage of depression. After detection, we can suggest them go to a psychiatrist for better treatment of their mental disorder. Also, we can help to  prevent them from doing any life-threatening occurrence owing to mental disorders. This work is done with a very limited dataset. In future work, the dataset will be increased. Also, we will implement more traditional and deep learning models in order to predict depression from the text so that better comparison can be done.   

% \section*{Acknowledgment}
% \input{Tex_Code/6.acknowledge/acknowledgement}    
\printbibliography

@misc{ tf_idf_link,
  author = {{Luthfi Ramadhan}},
  title = "TF-IDF Simplified",
%   year = "2020",
   howpublished = {\url{https://towardsdatascience.com/tf-idf-simplified-aba19d5f5530}}, 
  note = "[Online: accessed 30-October-2022]"
}

@article{svoboda_brychcín,
  author  = {Svoboda, L. and  Brychcín, T.},
  title   = {Enriching Word Embeddings with Global Information and Testing on Highly Inflected Language},
  journal = {Computación y Sistemas},
  year    = {2019},
  volume  = {23},
  number  = {03},
  pages   = {773-783},
  url     = {https://www.cys.cic.ipn.mx/ojs/index.php/CyS/article/view/3268}
}

@InProceedings{mikolov_zweig,
  author={Mikolov, T. and Yih, W. and Zweig, G.},
  title ={ Linguistic regularities in continuous space word representations}, 
  booktitle={ Human Language
Technologies Conference of the North American Chapter of the Association of Computational Linguistics(NAACL-HLT), ; Atlanta, GA, USA, June 9-14 }, 
  year=2013
 }

@article{martindale_mcKenzie,
  author  = {Martindale, C. and McKenzie, D.},
  title   = {On the utility of content analysis in author attribution:The Federalist},
  journal = {Computers and the Humanities},
  year    = {1995},
  volume  = {29},
  number  = {04},
  pages   = {259–270},
  url     = {https://link.springer.com/article/10.1007/BF01830395}
}

@InProceedings{abbasi_chen2,
  author={Abbasi, A. and Chen, H.},
  title ={Applying Authorship Analysis to Arabic Web Content}, 
  booktitle={Intelligence and Security Informatics, Berlin, Heidelberg}, 
  year=2005
 }

@InProceedings{Otoom_Amer,
  author={Otoom, F. A. and Abdullah, E. E. and Jaafer, S. and Hamdallh, A. and Amer, D.},
  title ={Towards author identification of Arabic text articles}, 
  booktitle={International Conference on Information and Communication Systems (ICICS), Irbid, Jordan, April 1-3}, 
  year=2014
 }

@article{JMLR_speed_gpu,
  author  = {Alexander G. de G. Matthews and Mark van der Wilk and Tom Nickson and Keisuke Fujii and Alexis Boukouvalas and Pablo Le{\'o}n-Villagr{\'a} and Zoubin Ghahramani and James Hensman},
  title   = {GPflow: A Gaussian Process Library using TensorFlow},
  journal = {Journal of Machine Learning Research},
  year    = {2017},
  volume  = {18},
  number  = {40},
  pages   = {1-6},
  url     = {http://jmlr.org/papers/v18/16-537.html}
}

@misc{shipra_saxena,
  author = {{Shipra Saxena}},
  title = "Binary Cross Entropy/Log Loss for Binary Classification", 
   howpublished = {\url{https://www.analyticsvidhya.com/blog/2021/03/binary-cross-entropy-log-loss-for-binary-classification/}}, 
  note = "[MARCH 3, 2021]"
}

@InProceedings{sharma_ochin,
  author={ Sharma, O.},
  title ={A New Activation Function for Deep Neural Network, Faridabad, India}, 
  booktitle={International Conference on Machine Learning, Big Data, Cloud and Parallel Computing (COMITCon),  Feb 14-16}, 
  year=2019
 }

@article{Kusumawati_2019,
  title={Comparison Performance of Naive Bayes Classifier and Support Vector Machine Algorithm for Twitter’s Classification of Tokopedia Services},
  author={Kusumawati, R. and  D'arofah, A. and Pramana, P. A.},
  journal={Journal of Physics Conference Series},
  volume={1320},
  pages={012016},
  year={2019}
}

@misc{ git_stop_word_bangla,
  author = {{stopwords-iso}},
  title = "Stopwords Bengali", 
   howpublished = {\url{https://github.com/stopwords-iso/stopwords-bn}}, 
  note = "[Online: accessed 31-May-2021]"
}

@misc{ spoken_language,
  author = {{Bengali language}},
  title = "Bengali language --- {W}ikipedia{,} The Free Encyclopedia",
%   year = "2020",
   howpublished = {\url{https://en.wikipedia.org/wiki/Bengali_language}}, 
  note = "[Online: accessed 04-February-2020]"
}

@article{abbasi_chen,
  title={Applying authorship analysis to extremist-group Web forum messages},
  author={Abbasi, A. and Chen, H.},
  journal={IEEE Intelligent Systems},
  volume={20},
  number={5},
  pages={67-75},
  year={2005}
}

@InProceedings{mikolov_kai,
  author={ Mikolov, T. and Chen, K. and Corrado, G. and Dean, J.},
  booktitle ={arXiv preprint arXiv:1301.3781}, 
  title={Efficient Estimation of Word Representations in Vector Space}, 
  year=2013
 }

@InProceedings{tripto_ali,
  author={Tripto, N. I. and Ali, M. E.},
  title ={Detecting Multilabel Sentiment and Emotions from Bangla YouTube Comments, Sylhet, Bangladesh}, 
  booktitle={International Conference on Bangla Speech and Language Processing (ICBSLP), Sept 21-22}, 
  year=2018
 }

@InProceedings{corney_anderson,
    author = {Corney, M. and Vel, O. de and Anderson, A. and Mohay, G.},
    title = {Gender-preferential text mining of e-mail discourse},
    booktitle = {Computer Security Applications Conference(CSAC), Las Vegas, USA, Dec 9-13},
    year = 2002
}

@article{shalabi_kanaanbt,
  title={Author gender identification from Arabic text},
  author={Alsmearat, K. and Al-Ayyouba, M. and Al-Shalabi, R. and Kanaanbt, G.},
  journal={Journal of Information Security and Applications},
  volume={35},
  number={8},
  pages={85-95},
  year={2017}
}

@article{Bsir_zrigui,
  title={Enhancing Deep Learning Gender Identification with Gated Recurrent Units Architecture in Social Text},
  author={Bsir, B. and Zrigui,M.},
  journal={Computación y Sistemas},
  volume={22},
  number={3},
  pages={757-766},
  year={2018}
}

@article{Rafiqul_kabir_Raihan,
  title={Depression detection from social network data using machine learning techniques},
  author={Islam, Md Rafiqul and Kabir, Muhammad Ashad and Ahmed, Ashir and Kamal, Abu Raihan M and Wang, Hua and Ulhaq, Anwaar},
  journal={Health information science and systems},
  volume={6},
  pages={1--12},
  year={2018},
  publisher={Springer}
}

@inproceedings{hassan_Jamil_2017sentiment,
  title={Sentiment analysis of social networking sites (SNS) data using machine learning approach for the measurement of depression},
  author={Hassan, Anees Ul and Hussain, Jamil and Hussain, Musarrat and Sadiq, Muhammad and Lee, Sungyoung},
  booktitle={2017 international conference on information and communication technology convergence (ICTC)},
  pages={138--140},
  year={2017},
  organization={IEEE}
}

@article{billah2019depression,
  title={Depression detection from Bangla Facebook status using machine learning approach},
  author={Billah, Masum and Hassan, Enamul},
  journal={Int. J. Comput. Appl},
  volume={975},
  pages={8887},
  year={2019}
}

@inproceedings{wang2013depression,
  title={A depression detection model based on sentiment analysis in micro-blog social network},
  author={Wang, Xinyu and Zhang, Chunhong and Ji, Yang and Sun, Li and Wu, Leijia and Bao, Zhana},
  booktitle={Trends and Applications in Knowledge Discovery and Data Mining: PAKDD 2013 International Workshops: DMApps, DANTH, QIMIE, BDM, CDA, CloudSD, Gold Coast, QLD, Australia, April 14-17, 2013, Revised Selected Papers 17},
  pages={201--213},
  year={2013},
  organization={Springer}
}

@InProceedings{Rambocas_Gama,
author = {M.Rambocas, and J. Gama,},
    title = {Marketing Research: The Role of Sentiment
Analysis},
    booktitle = {The 5th SNA-KDD Workshop11. University of Porto},
    year = 2013
}

@inproceedings{hassan2017sentiment,
  title={Sentiment analysis of social networking sites (SNS) data using machine learning approach for the measurement of depression},
  author={Hassan, Anees Ul and Hussain, Jamil and Hussain, Musarrat and Sadiq, Muhammad and Lee, Sungyoung},
  booktitle={2017 international conference on information and communication technology convergence (ICTC)},
  pages={138--140},
  year={2017},
  organization={IEEE}
}

@inproceedings{uddin2019depression,
  title={Depression analysis of bangla social media data using gated recurrent neural network},
  author={Uddin, Abdul Hasib and Bapery, Durjoy and Arif, Abu Shamim Mohammad},
  booktitle={2019 1st International Conference on Advances in Science, Engineering and Robotics Technology (ICASERT)},
  pages={1-6},
  year={2019},
  organization={IEEE}
}

@INPROCEEDINGS{abdul_hasib_arif,
  author={Uddin, Abdul Hasib and Bapery, Durjoy and Arif, Abu Shamim Mohammad},
  booktitle={2019 International Conference on Computer, Communication, Chemical, Materials and Electronic Engineering (IC4ME2)}, 
  title={Depression Analysis from Social Media Data in Bangla Language using Long Short Term Memory (LSTM) Recurrent Neural Network Technique}, 
  year={2019},
  volume={},
  number={},
  pages={1-4},
  doi={10.1109/IC4ME247184.2019.9036528}}

\vspace{12pt}
% \color{red}
% IEEE conference templates contain guidance text for composing and formatting conference papers. Please ensure that all template text is removed from your conference paper prior to submission to the conference. Failure to remove the template text from your paper may result in your paper not being published.

\end{document}